\documentclass[epj]{svjour}
\usepackage{graphics}
\begin{document}
\title{Recent results from parton cascade/microscopic transport}
\author{Bin Zhang
}                     
%
%
\institute{Department of Chemistry and Physics,
Arkansas State University, State University, AR 72467-0419, USA}
\date{Received: date / Revised version: date}
%
\abstract{
Parton cascade is a microscopic transport approach 
for the study of the space-time evolution of the 
Quark-Gluon Plasma produced in relativistic heavy 
ion collisions and its experimental manifestations. 
In the following, parton cascade calculations on 
elliptic flow and thermalization will be discussed.
Dynamical evolution is shown to be important
for the production of elliptic flow including
the scaling and the breaking of the scaling of
elliptic flow. The degree of thermalization is
estimated using both an elastic parton cascade
and a radiative transport model. A longitudinal
to transverse pressure ratio, $P_L/P_T\approx 0.8$,
is shown to be expected in the central cell
in central collisions. This provides
information on viscous corrections to the ideal
hydrodynamical approach.
} 
\maketitle
\section{Introduction}
\label{sec:intro}

Quantum Chromodynamics (QCD) is the theory of strong interactions.
To understand strong interaction phenomena, people invented many
QCD motivated models. One such model is the parton cascade model.
It is a natural continuation of the hadron transport 
approach \cite{Pang:1992sk,Li:1995pra,Sorge:1989vt,Bass:1998ca}. 
One major advantage of this microscopic
transport description of relativistic heavy ion collisions
is that it does not rely on the assumption of local thermal
equilibrium. In other words, it can be used to study the
equilibration process. The parton cascade concept was 
introduced by Klaus Kinder-Geiger and Berndt M\"uller 
\cite{Geiger:1990bu,Geiger:1991nj}. 
Klaus named his parton cascade program \textsc{VNI}
(``Vincent Le CuCurullo Con GiGinello" for ``that little 
guy who plays with quarks and gluons") \cite{Geiger:1997pf}. 
Many developments of \textsc{VNI} and other parton cascades 
\cite{Bass:2002fh,Zhang:1997ej,Molnar:2001ux,Zhang:1999bd,Xu:2004mz}
have lead to new insights
into Quark-Gluon Plasma production. 

Parton cascade describes the evolution of the partonic system by 
solving the Boltzmann equation. Some simplifications are 
introduced to make the problem tractable. Interactions
between particles and the color field are not included. 
Particles' color degrees of freedom are not followed. 
Under these conditions, the Boltzmann equation can be written as
\begin{eqnarray*}
\lefteqn{\left(\frac{\partial}{\partial t}+\frac{\vec{p}}{E}
\cdot
\frac{\partial}{\partial\vec{x}}\right)
f(\vec{x},\vec{p},t) = } \hspace{0.8in}   \\ 
&\mbox{     } & S(\vec{x},\vec{p},t)+C_{22}+C_{23}+C_{32}+\cdots .
\end{eqnarray*}

In the above equation, $f(\vec{x},\vec{p},t)$ is the phase-space 
distribution. Its time evolution depends on the source term 
$S(\vec{x},\vec{p},t)$ and the collision terms $C_{22}$, $C_{23}$, 
$C_{32}$, etc. The source term describes particle production 
from processes other than direct collisions, e.g., particle
production from strong color field or Glasma \cite{Lappi:2006fp}. 
Each collision term, $C_{mn}$, 
describes collisions with $m$ incoming particles and $n$ outgoing 
particles. It is an integral of dimensions $3\times(m+n-1)-4$
with $m$ phase-space distributions. 
In general, the evolution of the phase-space distribution can not
be solved analytically. One way of solving the Boltzmann equation 
numerically is to discretize the phase space distribution. It can 
be written as a sum of contributions from point particles,
\begin{eqnarray*}
\lefteqn{f(\vec{x},\vec{p},t)=}\hspace{0.2in}\\
&&\sum_{i=1}^n w_i \delta^{(3)}\left( \vec{x}-
\left(\vec{x}_i-\frac{\vec{p}_i}{E_i}(t-t_i)\right)\right)
\delta^{(3)}(\vec{p}-\vec{p}_i) .
\end{eqnarray*}
Where particle $i$ with weight $w_i$ and four-momentum 
$(\vec{p}_i,E_i)$ propagates in a straight line from production 
position $\vec{x}_i$ at production time $t_i$. Then Monte Carlo 
method is used to evaluate the collision terms.

In the following, recent results from parton cascade and 
microscopic transport will be reported. The main focus 
will be on elliptic
flow (Sec.~\ref{sec:v2}) and thermalization 
(Sec.~\ref{sec:thermalization}). The former is important
for the understanding of experimental data and the latter offers
insight into the macroscopic description of heavy ion collisions.
The discussion will end with a summary and outlook.

\section{Elliptic flow}
\label{sec:v2}

Elliptic flow measures transverse momentum anisotropy relative
to the reaction plane. It can be 
characterized by the second Fourier coefficient of the particle 
azimuthal distribution. If a Quark-Gluon Plasma is produced, 
large elliptic flow is expected based on the fact that partons
form early than hadrons and the partonic equation of state is 
harder than the hadronic equation of state. Since the elliptic 
flow is 
produced early during the evolution, it reflects interactions 
of partons inside the Quark-Gluon Plasma. This was indeed
demonstrated by Zhang, Gyulassy and Ko \cite{Zhang:1999rs}.
As RHIC data came out \cite{Ackermann:2000tr}, 
a study using more realistic
diffuse transverse gluon geometry instead of overlapping
cylinders was carried out by Moln\'ar and Gyulassy 
\cite{Molnar:2001ux}. It was found out that large elastic 
cross sections (on the order of 45mb) are 
needed in order to describe RHIC elliptic flow data.
This certainly implies 
that processes other than perturbative elastic gluon scatterings 
can be important. Further calculations were carried out
by Xu and Greiner \cite{Xu:2004mz}. 
They implemented the two gluons 
to three gluons and its inverse reaction. The 
Landau-Pomeranchuk-Migdal (LPM) effect is approximated 
by cutting off 
radiated gluons that can not form before the next collision. 
This leads to large angle radiative scatterings which facilitate 
thermalization and the buildup of elliptic flow. Under the 
assumption of local parton-hadron duality, Xu, Greiner and
St\"ocker showed that the calculated 
elliptic flow matches experimental data if the strong
interaction coupling constant $\alpha_s=0.6$ 
is used \cite{Xu:2007jv}.

Within the elastic parton cascade framework, recent works have 
been done to investigate the relation between the macroscopic 
description and the microscopic description of 
heavy ion collisions \cite{Zhang:2003wa,Molnar:2008xj}. 
Moln\'ar and Huovinen showed that the microscopic and 
macroscopic descriptions 
agree in the description of transverse momentum $p_t$ 
differential flow \cite{Molnar:2008xj}. This is 
true not only for the case with a very large constant rescattering 
cross section, but also for the case with a time dependent cross 
section tuned such that the shear viscosity to entropy
density ratio $\eta/s$ is approximately $1/(4\pi)$. 
In other words, when the macroscopic description, i.e., viscous 
hydro applies, the macroscopic description is equivalent to the 
microscopic description in describing experimental observables. 
When radiative processes are included, $\eta/s$ is shown to be 
very small. In particular, if $\alpha_s=0.6$ is used, $\eta/s$ 
is close to $1/(4\pi)$ \cite{Xu:2007jv}. 
This is what motivated many recent works
on parton transport at minimal viscosity. One recent 
progress in this area is the demonstration of elliptic 
flow scaling \cite{Ferini:2008he} by Ferini, Colonna, Di Toro
and Greco. 
It is shown that when freeze-out happens at an energy
density value of $\epsilon=0.2$ 
GeV/fm$^3$, the $p_t$ differential elliptic flow scales with
both the initial spatial anisotropy $\epsilon_x$ and 
with the integrated elliptic flow $\langle v_2\rangle$. 
However, when the system freezes out at $\epsilon=0.5$ GeV/fm$^3$, 
the $v_2/\epsilon_x$ scaling is broken, while the 
$v_2/(k\langle v_2\rangle)$ still holds. This interesting
result revised the common belief that the integrated $v_2$ is 
a good measure of the initial spatial anisotropy. The
breaking of $v_2/\epsilon_x$ scaling is larger for larger
impact parameter. It is consistent with the breaking 
of $v_2/\epsilon_x$ scaling observed 
experimentally \cite{Alver:2006wh}. 
This breaking gives valuable information about the evolution and
freeze-out of the partonic system. 

\section{Thermalization}
\label{sec:thermalization}

Ideal hydrodynamics is very successful in describing RHIC data 
\cite{Teaney:2000cw,Huovinen:2001cy,Kolb:2001qz,Hirano:2005xf}. 
One assumption of ideal hydrodynamics is local thermal equilibrium.
Hydro dynamical equations can also be used when there is local 
isotropy \cite{Heinz:2005zi}. In this case, additional 
entropy production equation 
needs to be used in place of entropy conservation. Quantum 
Mechanics requires non-zero viscosity \cite{Danielewicz:1984ww}. 
The local equilibrium 
assumption of ideal hydrodynamics implies zero viscosity. Hence 
ideal hydrodynamics is only an effective description. 
The degree of 
thermalization is an important aspect of improving the ideal 
hydrodynamical description of relativistic heavy ion 
collisions \cite{Romatschke:2007mq,Song:2007ux,Dusling:2007gi}.

The evolution of bulk properties of the hot and dense
matter produced in relativistic heavy ion collisions
was studied recently in the frame work of 
the \textsc{AMPT} (A Multi-Phase Transport) model 
\cite{Zhang:2008zzk}. The \textsc{AMPT} model 
is a hybrid model
\cite{Zhang:1999bd,Lin:2000cx,Lin:2001yd,Lin:2004en}.
It uses \textsc{HIJING} 
(Heavy Ion Jet-INteraction Generator) 
\cite{Wang:1991hta} as
the initial condition. The publicly available version
has two options: the default model and the string melting
model. In the parton stage, the default model has only
mini-jet gluons while the string melting model melts
hadrons in HIJING into quarks and anti-quarks according
to their valence structures. The parton evolution is 
handled with the \textsc{ZPC} (Zhang's Parton Cascade) 
parton cascade model  \cite{Zhang:1997ej}. The default
model hadronizes via the Lund string fragmentation 
\cite{Sjostrand:1993yb} while
the string melting model uses a simple coalescence model
\cite{Greco:2003xt,Greco:2003mm}
for hadronization. The \textsc{ART} (A Relativistic Transport) 
model \cite{Li:1995pra,Li:2001xh} is used for the
hadron transport processes. The default model
is good at describing particle distributions 
\cite{Lin:2000cx,Lin:2001yd}
and the string melting model is needed for the 
description of elliptic flow \cite{Lin:2001zk}
and HBT radii \cite{Lin:2002gc}. Other observables have
also been studied using the \textsc{AMPT} model
\cite{Pal:2002aw,Chen:2004dv,Chen:2004vha,Chen:2005mr,Chen:2005zy,Zhang:2000nc,Zhang:2002ug,Zhang:2005ni,Chen:2006vc}.

\begin{figure}
\resizebox{0.45\textwidth}{!}{%
  \includegraphics{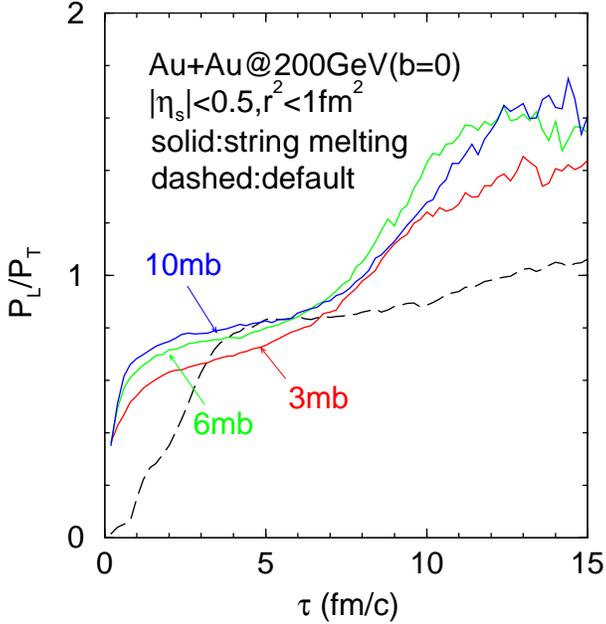}
}
\caption{Evolution of $P_L/P_T$ from the \textsc{AMPT} model.}
\label{fig:plopt_ampt1}       
\end{figure}

The bulk properties of matter
in the central cell in central collisions are studied. In 
particular, the equation of state is characterized by the pressure
over energy density ratio $P/\epsilon$ as a function of energy 
density. In the central cell, the equations of state of both the 
default model and the string melting model interpolate between the 
hard partonic phase and the soft hadronic phase. However, they 
differ in details, especially in the intermediate 
energy density range. 
It is important to notice that unlike for the equilibrium case, 
the equation of state is not the whole story. To gauge the degree 
of equilibration, additional information is needed. For central 
heavy ion collisions, because of cylindrical symmetry, the 
longitudinal pressure to transverse pressure ratio, $P_L/P_T$, 
can be used to characterize the degree of thermalization. 
More precisely, it is a measure of the degree of isotropization. 
Fig.~\ref{fig:plopt_ampt1} shows $P_L/P_T$ from 
the \textsc{AMPT} model. Due to 
the increase of parton production in the initial stage, the string 
melting model has a faster increase in the early stage. As the 
parton cross section increases, $P_L/P_T$ goes closer to 1. 
The pressure anisotropy approaches 1, then it increases and 
crosses 1. In this particular case, $P_L/P_T=1$ does not imply 
thermalization. It is caused by the onset of transverse 
expansion. This also shows up in the change of
slope in the energy density evolution. 
The exponent of energy density evolution changes as transverse 
expansion sets in. This certainly demonstrates that only partial
thermalization is achieved.

%

\begin{figure}
\resizebox{0.45\textwidth}{!}{%
  \includegraphics{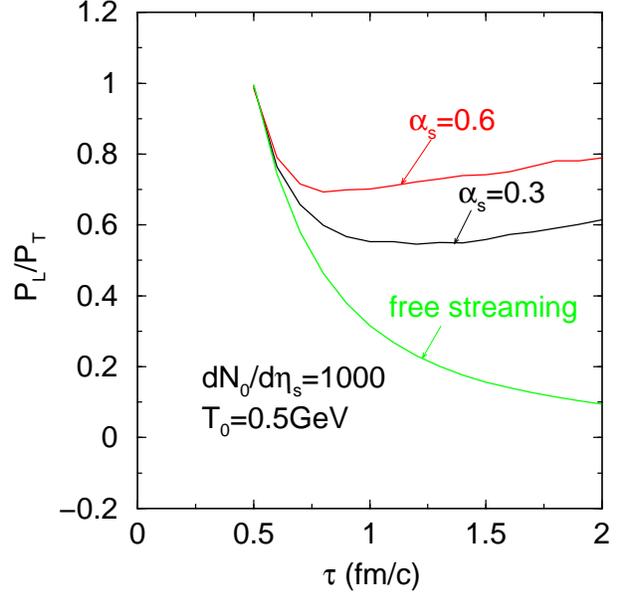}
}
\caption{$P_L/P_T$ evolution starting with a local
thermal initial condition.}
\label{fig:plopt_rbs1}       
\end{figure}

To study more carefully the effects of radiative processes, 
we will look at whether pressure isotropy can be maintained 
by gluon elastic and inelastic processes. For this purpose, 
we start with an initial local thermal gluon distribution.
Gluons are produced at a proper time $\tau=0.5$ fm/c. The 
initial space-time rapidity distribution is uniform with a 
space-time rapidity density of 1000 and the gluons are 
distributed between space-time rapidity $\eta_s=-5$ and +5.
In the transverse direction, at the formation proper 
time, they 
are distributed uniformly within a radius of 5 fm. We will 
start with an initial temperature $T_0=0.5$ GeV. The
evolution in $\tau$ is studied with fixed grid in $\eta_s$
and expanding grid in the transverse direction. 16 test
particles per real particle are used in the calculations.
Fig.~\ref{fig:plopt_rbs1} gives the time evolution of $P_L/P_T$. 
One can use the free streaming curve as a reference for 
the effect of expansion. In the free streaming case, if 
the average longitudinal momentum squared over
the average transverse momentum squared, $<p_l^2>/<p_t^2>$, 
is used to measure the isotropy, the 
evolution follows exactly $1/\tau^2$. For the $P_L/P_T$ 
evolution, free streaming can be approximated by 
$1/\tau^2$, but more precisely, it evolves slightly slower 
than $1/\tau^2$. Now we turn on 2 to 2 (two incoming gluons 
and two outgoing gluons), 2 to 3 and its inverse reaction. 
The cross sections are chosen to be isotropic to maximize 
equilibration. $\sigma_{23}$ is fixed to be 20\% of 
$\sigma_{22}$ on the same order as measured 
in Ref.~\cite{Xu:2004mz}. $I_{32}$ is 
determined by detailed balance. $\sigma_{22}$ is taken to 
be the screened Coulomb cross section with the screening 
mass determined dynamically. Two values of the strong 
interaction coupling constant $\alpha_s$ are used. The 
figure shows that when interactions are turned on, 
$P_L/P_T$ evolution deviates from the free streaming curve. 
The longitudinal to transverse pressure ratio 
becomes larger as $\alpha_s$ increases. Due to 
the competition of expansion and thermalization, there is a 
minimum in the $P_L/P_T$ evolution. The larger the strong 
coupling constant is, the stronger the thermalization is, 
and the earlier the time that the minimum occurs.

%

\begin{figure}
\resizebox{0.45\textwidth}{!}{%
  \includegraphics{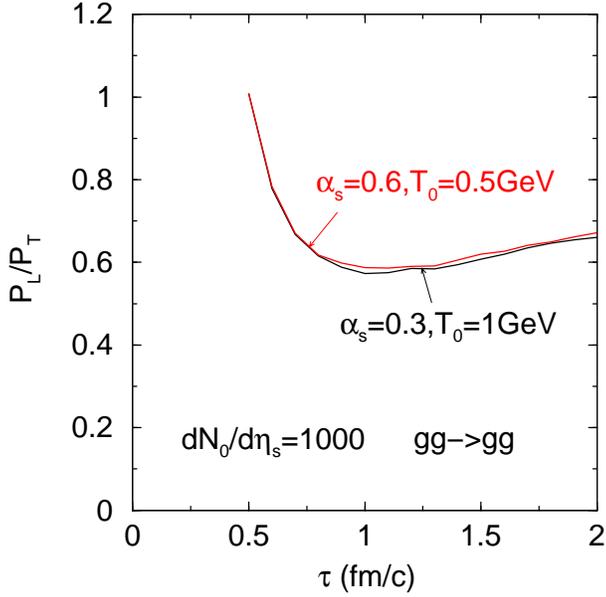}
}
\caption{$\alpha_s T_0$ scaling in $P_L/P_T$ evolution for elastic
scattering with dynamical screening mass.}
\label{fig:plopt_rbs2}       
\end{figure}

\begin{figure}
\resizebox{0.45\textwidth}{!}{%
  \includegraphics{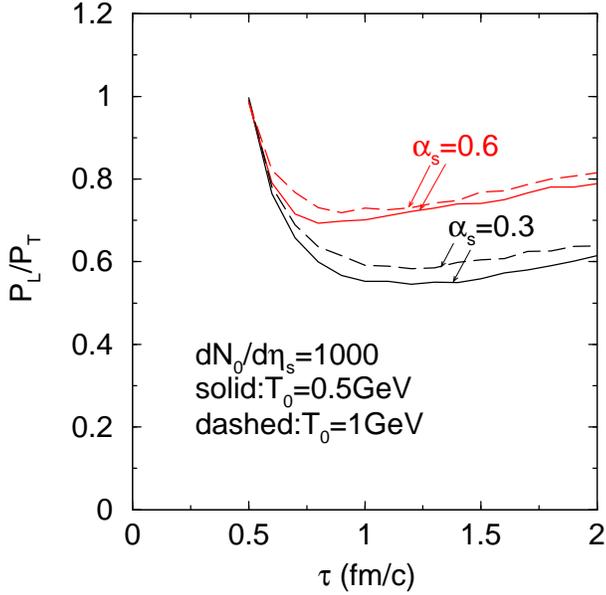}
}
\caption{Approximate $\alpha_s$ scaling in $P_L/P_T$ 
evolution.}
\label{fig:plopt_rbs3}       
\end{figure}

It is interesting to see what happens if the parameters
of the system change. It turns out there is a scaling
law for the case with elastic collisions with dynamical
screening. More precisely, the $P_L/P_T$ evolution 
depends only on the combination $\alpha_s T_0$. This is
illustrated in Fig.~\ref{fig:plopt_rbs2}. The two curves in
Fig.~\ref{fig:plopt_rbs2} have different initial thermal
conditions. The solid one has $T_0=0.5$ GeV and the
dashed one has $T_0=1$ GeV. However, they have the
same $\alpha_s T_0=0.3$ GeV. The $P_L/P_T$ curves agree
with each other. This scaling is caused by
the same initial density ($n_0$) and the same initial 
binary cross section ($\sigma_{22}\propto \alpha_s T_0/n_0$).
This combination leads to the same evolution of 
density and pressure isotropy. When 2 to 3 and 3 to 2
are turned on, different systems with different
initial temperatures will evolve toward different
chemical equilibrium densities. This breaks the
$\alpha_s T_0$ scaling. Fig.~\ref{fig:plopt_rbs3} shows
the evolution of $P_L/P_T$ when 2 to 3 and 3 to 2
are included. It is interesting to notice that
though the $\alpha_s T_0$ scaling is broken, the
evolution has an approximate $\alpha_s$ scaling. This
certainly demonstrates the important of particle
number changing processes in thermalization. The
case with $\alpha_s=0.6$ gives a value of $P_L/P_T$
that is around $0.8$. It is consistent with
the \textsc{AMPT} results with parton rescattering
cross section $\sigma=10$ mb as shown in 
Fig.~\ref{fig:plopt_ampt1}. Calculations from other
models \cite{Bravina:2000dk,Xu:2007aa,Huovinen:2008te}
also give comparable anisotropy.
This level of anisotropy should be expected
in the improved viscous hydrodynamical
studies.

%

%

\section{Summary and outlook}
\label{sec:summary}

The parton cascade/microscopic transport method has
been used to study relativistic heavy ion collisions.
It has made important contributions to the understanding
of the physics behind elliptic flow and thermalization.
In particular, transport theory and viscous hydrodynamics
agree for the description of dense systems. Parton 
freezeout is seen to be important for elliptic flow 
scaling. The degree of kinetic equilibration depends
on detailed transport processes. Radiative processes
can have unique contributions to thermalization.

A recent study by Fochler, Xu and Greiner 
\cite{Fochler:2008ts} 
shows that parton transport may be
able to give a unified description of
 both jet quenching and the large $p_t$ differential
elliptic flow. This is not easy by straightforward
application of jet quenching formalism as the
expansion of the system makes it difficult to
generate enough elliptic 
flow \cite{Gyulassy:2001kr}. Straightforward
application of jet quenching formalism also faces
problem with explaining both the light quark
jet quenching and the heavy quark jet quenching.
Several studies have shown the importance of
transport processes, including approaches with
heavy resonances \cite{vanHees:2004gq,vanHees:2005wb}, 
the Langevin model approach \cite{Moore:2004tg},
effective large elastic cross sections 
for heavy quarks \cite{Zhang:2005ni}, 
early formation and dissociation of 
heavy mesons \cite{Adil:2006ra},
and the enhancement of $\Lambda_c/D$ ratio
in heavy ion collisions \cite{Sorensen:2005sm}.
Partonic transport can certainly
contribute to the understanding of this phenomenon
by including these mechanisms into the space-time
evolution of the system. There are also recent
parton transport studies on the ridge and 
Mach cone phenomena \cite{Ma:2006fm,Ma:2006rn,Ma:2008nd}
that show the importance of strong interactions. 
Further studies with partonic transport
will help to quantitatively understand these
and many other phenomena.

\section*{Acknowledgments}
B.Z. thanks L.W. Chen and C.M. Ko for collaboration on
the \textsc{AMPT} project and S.A. Bass, U. Heinz and T. Renk 
for helpful discussions. We also thank the Parallel 
Distributed Systems Facilities of the National 
Energy Research Scientific Computing Center 
for providing computing resources. This work was supported 
by the U.S. National Science Foundation under 
Grant No. PHY-0554930.



%

\end{document}